\documentclass[prb,twocolumn,showpacs,preprintnumbers,amsmath,amssymb]{revtex4-1}

\usepackage{graphicx}
\usepackage{dcolumn}
\usepackage{bm}

\begin{document}

\title{Multiband Te $p$ Based Superconductivity of Ta$_4$Pd$_3$Te$_{16}$}

\author{David J. Singh}

\affiliation{Materials Science and Technology Division,
Oak Ridge National Laboratory, Oak Ridge, Tennessee 37831-6056}

\date{\today}

\begin{abstract}
Ta$_4$Pd$_3$Te$_{16}$ is a recently discovered superconductor that
has been suggested to be an unconventional superconductor near
magnetism. We report electronic structure calculations showing that
in spite of the layered crystal structure the material is an 
anisotropic three dimensional metal.
The
Fermi surface contains prominent one and two dimensional features, including
nested 1D sheets, a 2D cylindrical
section and a 3D sheet.
The electronic states that make up the Fermi surface are mostly
derived from Te $p$ states, with small Ta $d$ and Pd $d$ contributions.
This places the compound far from magnetic instabilities. 
The results are discussed in terms of multiband superconductivity.
\end{abstract}


\maketitle

\section{introduction}

Jiao and co-workers recently discovered
superconductivity with a critical temperature $T_c$=4.6 K
in Ta$_4$Pd$_3$Te$_{16}$. \cite{jiao}
These authors concluded that there is evidence for strong electron-electron
interactions in this compound based on an analysis of the specific heat.
Moreover,
Pan and co-workers reported that
$T_c$ increases to over 6 K under pressure
and found a linear contribution to the temperature
dependent thermal conductivity, which
increases with magnetic field similar to cuprate superconductors. \cite{pan}
This implies an electronic thermal conductivity contribution in
the superconducting phase, i.e. line nodes or some part of the Fermi surface
that is not gapped.
These measurements, which extend to $T$=80 mK,
which is well below $T_c$, were taken
as implying an unconventional superconducting
state, perhaps associated with the quasi-one-dimensional features of the
crystal structure. \cite{pan}

The crystal structure (Fig. \ref{struct}) shows a layered structure 
with Ta and Pd atoms occurring in one dimensional chains. The structure
differs from those of the superconducting chalcogenides
Ta$_2$PdS$_5$, Nb$_2$Pd$_{0.8}$S$_5$ and Nb$_3$Pd$_{0.7}$Se$_7$
in being a stoichiometric compound with nominally flat two dimensional sheets
in the crystal structure.
\cite{zhang,zhang2,lu,singh-ta2}
We note that there is the possibility
that the layered crystal structure may be misleading from the point
of view of the electronic properties, as
previous electronic structure calculations by Alemany and
co-workers showed important interlayer Te -- Te interactions.
\cite{alemany}
In addition, it should be noted that there are considerable local
distortions around the different Te atoms in the structure.
Here we report electronic structure calculations in relation to
the above observations.

\section{methods}

\begin{figure}
\includegraphics[width=\columnwidth,angle=0]{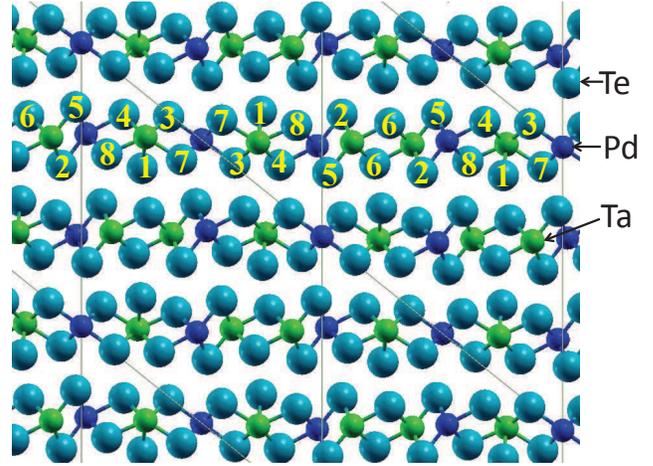}
\caption{(color online) Crystal structure of Ta$_4$Pd$_3$Te$_{16}$
viewed along the $c$-axis (in spacegroup 12,
$C2/m$, setting 2, unique axis $c$). The unit cell is indicated by the lines.
The atomic positions are from the total energy minimization.
The Te atoms in one layer are labeled according to the inequivalent
Te positions in the structure (see text).}
\label{struct}
\end{figure}

The calculations reported here were performed using standard
density functional theory with the generalized gradient approximation
of Perdew, Burke and Ernzerhof (PBE-GGA). \cite{pbe}
We used the general potential linearized augmented planewave (LAPW)
method \cite{singh-book} as implemented in the WIEN2k code. \cite{wien2k}
We used well converged LAPW basis sets plus local orbitals to treat
the semicore states. The LAPW sphere radii were $R$=2.5 bohr for all atoms,
and the planewave sector cutoff, $k_{max}$ was set according to $R k_{max}$=9.

The monoclinic
lattice parameters were taken from the experimental data of Mar and
Ibers, \cite{mar}
$a$=21.2762 \AA, $b$=19.510 \AA, $c$=3.735 \AA, $\gamma$=128.825$^\circ$
(note we show results using spacegroup 12, $C2/m$ setting 2, unique axis $c$,
while the original paper of Mar and Ibers uses setting 1, unique axis $b$).
The internal atomic coordinates were then determined by
total energy minimization with the PBE-GGA starting from the experimental
structure including relativity at the scalar relativistic level
for the valence bands. The resulting structure, shown in
Fig. \ref{struct}, was used for the electronic
structure calculations. These were performed self-consistently including
spin-orbit for all states.
The shortest Te -- Te distance that couples different layers in the
relaxed structure is 3.61 \AA, which is short enough for significant
interlayer Te -- Te interactions, as were discussed by Alemany and co-workers.
\cite{alemany}

The relaxed structure (Fig. \ref{struct}) 
is quite complex as seen.
In particular, while nominally the structure could be described as Te
bilayers with metal atoms in the interstitial sites, it is clear that
these bilayers are very strongly distorted. These distortions lead
to interesting structural features, such as short bonded Te zig-zag chains
along the $c$-axis composed of the Te1 and Te8 atoms, as labeled in the
figure, with a Te-Te bond length of only 3.22 \AA.
Also, there are large differences between the different Te sites.
The nearest Te-Te distances for the
different Te atoms run from 3.08 \AA, for the Te3-Te4 bond,
to 3.73 \AA, for Te2. Also the bond valence sums are far from the
nominal value and vary among the different Te, ranging from
2.68 (Te7) to 2.91 (Te8). This implies an important role for Te-Te
bonding.
The experimentally reported crystal structure refinement \cite{mar} shows
similar features including both the range of Te-Te bond lengths
and the variation in bond valence sums.
While this is unusual, it is reminiscent of IrTe$_2$,
which shows a first order structural transition related to
frustrated Te-Te $p$ bonding.
\cite{fang,cao}

\section{electronic structure}

\begin{figure}
\includegraphics[height=\columnwidth,angle=270]{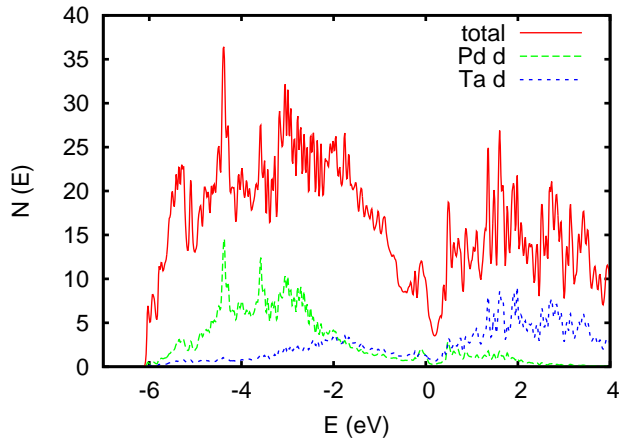}
\caption{(color online) Calculated electronic density of states
and projections of Pd $d$ and Ta $d$ character on a per formula
unit basis. The Fermi energy is at 0 eV.}
\label{dos}
\end{figure}

The calculated electronic density of states and projections of
Pd $d$ and Ta $d$ character are shown in Fig. \ref{dos}.
There is strong hybridization between both types of metal atoms
and the Te evident in this plot. The Ta and Pd $d$
contributions occur in broad peaks above and below the
Fermi level, $E_F$. As seen, the Pd $d$ orbitals are
mostly occupied, while the Ta $d$ orbitals are mostly unoccupied.
In a fully ionic picture, one might assign valences Ta$^{+5}$,
Pd$^{+4}$ in order to obtain the correct stoichiometry with Te$^{-2}$
However Pd$^{4+}$, would seem to be a highly unlikely valence especially
in a telluride.
The Pd density of states with its nearly fully occupied
$d$ orbitals below $E_F$ is clearly not consistent with
a Pd$^{4+}$ state. Therefore important Te-Te p bonding
is expected, consistent with that seen in the density of states.

The calculated value at the Fermi energy is $N(E_F)$=9.6 eV$^{-1}$
per formula unit, which is significantly higher than the value of
5.5 eV$^{-1}$ obtained by Alemany and co-workers. \cite{alemany}
We did calculations for the unrelaxed atomic coordinates
\cite{mar} as well.
However, and find a value similar to but slightly higher
than the value with the relaxed structure.
All results shown here are with the relaxed structure.
The bare linear specific heat coefficient from our calculation is
22.7 mJ/(mol K$^2$). Comparing with the experimental value
of 42.8 mJ/(mol K$^2$), one infers an enhancement of 
(1+$\lambda_{tot}$)=1.89 or $\lambda_{tot}$=0.89,
i.e. an intermediate coupling
value. Importantly, as seen from the projections of the density of states,
there is very little transition metal contribution to $N(E_F)$, which
instead derives from Te $p$ states.
The Pd $d$ contribution to $N(E_F)$ based on projection onto the Pd
LAPW spheres is 1.3 eV$^{-1}$ summed over the three Pd atoms, while the
Ta contribution is similar at 1.3 eV$^{-1}$ summed over the four Ta atoms.
This strongly argues against
nearness to magnetism in this compound.

\begin{figure}
\includegraphics[width=\columnwidth,angle=0]{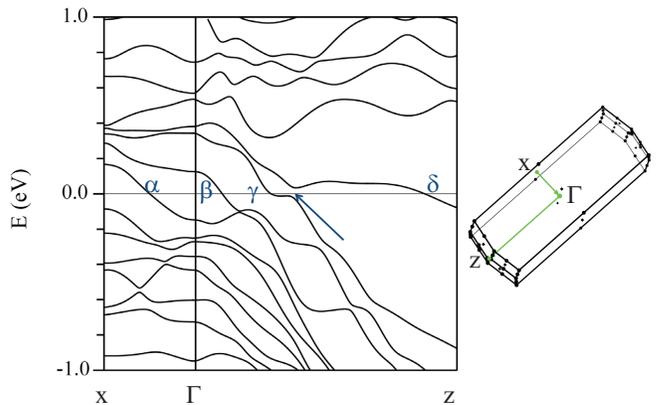}
\caption{(color online) Band structure of Ta$_4$Pd$_3$Te$_{16}$
along the lines shown in the right panel. The band crossings
corresponding to the four sheets of Fermi surface are labeled with
Greek letters. The arrow denotes the anticrossing as discussed in the
text.  The Fermi energy is at 0 eV.}
\label{bands}
\end{figure}

\begin{figure}
\includegraphics[width=\columnwidth,angle=0]{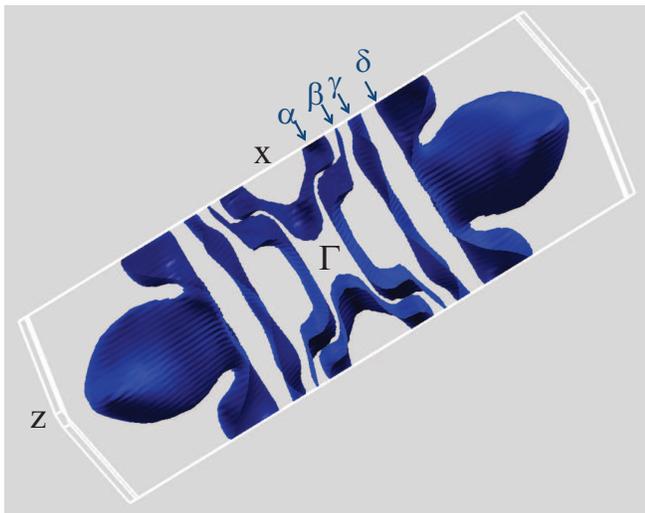}
\caption{(color online) Fermi surface Ta$_4$Pd$_3$Te$_{16}$.
The four sheets are labeled corresponding to the bands in Fig. \ref{bands}.}
\label{fermi}
\end{figure}

The band structure in the range around the Fermi energy, $E_F$ is given in
Fig. \ref{bands}. As shown, there are four bands crossing $E_F$,
labeled in the plot. These give rise to four sheets of
Fermi surface as depicted in Fig. \ref{fermi}.
The are a 2D hole cylinder (``$\alpha$"), two very 1D and
therefore nested sheets (``$\beta$" and ``$\gamma$")),
which may lead to nearness to
density wave instabilities, and a heavy 3D sheet (``$\delta$").
The 2D cylinder, $\alpha$, and the first 1D sheet, $\beta$ anticross making
the shapes of these surfaces more complex.
Nonetheless the nesting is at $\sim$ 0.1 $2\pi/c$
(from $\beta$) and $\sim$ 0.3 $2\pi/c$ (from $\gamma$),
i.e. along the direction of the metal chains.
The contributions to $N(E_F)$ from these are
1.4 eV$^{-1}$, 1.8 eV$^{-1}$, 1.0 eV$^{-1}$ and 5.4 eV$^{-1}$,
for sheets $\alpha$, $\beta$, $\gamma$ and $\delta$, respectively,
following the
labels of Fig. \ref{bands}.
In addition, there is an anticrossing with another Te
$p$ derived band very close to the Fermi energy
crossing of $\gamma$ as indicated by the
arrow in Fig. \ref{bands}.
This would come to the Fermi energy for very small levels of hole doping.
Such hole doping could come from Ta deficiency.
We emphasize that as is evident from the density of states, all of these
sheets of Fermi surface are derived from Te $p$ states and there is
little transition element $d$ character at the Fermi energy.
This implies that any nearby density wave instability due to nesting would be
a charge density wave (CDW) and not a spin density wave (SDW).
This density wave,
if it actually condensed, would consist of a structural modulation along
the short $c$-axis, metal chain direction, which should be clearly seen
in diffraction.

The mixture of 1D-like, 2D-like and 3D Fermi surfaces leads to
a net anisotropic, but 3D metal. The conductivity anisotropy was
estimated by integrating the transport function $\sigma/\tau$,
where $\tau$ is the scattering rate, taken as constant for all bands
and directions. The eigenvalues of the $\sigma/\tau$ tensor are
3.9 x 10$^{19}$ ($\Omega$ m s)$^{-1}$,
1.5 x 10$^{19}$ ($\Omega$ m s)$^{-1}$ and
10.5 x 10$^{19}$ ($\Omega$ m s)$^{-1}$,
with directions approximately
(note the monoclinic symmetry) in the layers, perpendicular to $c$,
approximately across the layers, and along $c$, respectively.
As noted, the nesting feature is along the $c$-axis direction.
This means that if a density wave condensed, or if there were strong
scattering associated with soft phonons related to the nesting, this
would primarily affect the nested sheets, and therefore the $c$-axis
direction conductivity.

We now turn to the details of the band structure. Because of the
low symmetry, the different bands have mixed orbital and atomic
character. The following is a description of the main characters.
The 2D $\alpha$ sheet is derived from a mixture of different Te $p$
states, and of the four sheets is the sheet with the most Ta $d$ character.
The 1D $\beta$ as mainly Te7 (Fig. \ref{struct}) character. This is
similar to the other 1D ($\gamma$) sheet, which in addition involves
Te1 and Te5 orbitals. The 3D, $\delta$ sheet, which is the sheet that
makes the largest contribution to $N(E_F)$ is mainly from Te3
$p$-orbitals, which are hybridized with Pd $d$ states.
Thus the different sheets of Fermi surface, and in particular
the nested quasi-1D sheets and the other sheets are associated
with different orbitals on different atoms.

\section{Discussion and Conclusions}

As mentioned, Ta$_4$Pd$_3$Te$_{16}$ is a low temperature superconductor,
with experimental signatures of possible unconventional behavior.
One such signature is an inconsistency between the
superconducting $\lambda_{sc}$ inferred from the specific heat jump,
which implies weak coupling behavior, and that inferred from the
enhancement of the low temperature specific heat,
$\gamma=\gamma_{bare}(1+\lambda_{tot})$. \cite{jiao}
In particular, Jiao and co-workers measured the specific heat jump as
$\Delta C/(\gamma T)$=1.4 $\pm$ 0.2 consistent with the weak coupling
BCS value of 1.43, implying low $\lambda_{sc}$.
This included a careful correction for the superconducting volume fraction.
They also inferred $\lambda_{tot}$=2.3 by combining the measured specific
heat $\gamma$ with density of states $N(E_F)$=5.5 eV$^{-1}$ from
the report of Alemany and co-workers. \cite{alemany}
However, we obtain a higher $N(E_F)$=9.6 eV$^{-1}$, which then gives a
correspondingly lower renormalization and $\lambda_{tot}$=0.89. This is
quite reasonable considering the specific heat jump. Therefore it
may not be necessary to consider an enhancement to $\lambda_{tot}$
due to non-phonon channels such as spin fluctuations to reconcile
the $\gamma$ with the specific heat jump.

We note that the transition metal $d$ contribution to $N(E_F)$
is low, which argues against spin-fluctuations.
Also, it should be noted that the Fermi surface consists of four
distinct sheets with contributions from different atoms.
Therefore it is reasonable to expect that the pairing interactions
could be stronger on some sheets than others, which would complicate
the simple interpretation of the specific heat jump using a single
band BCS formula.

From the point of view of superconductivity, nearness to
a CDW can provide an attractive interaction through the electron
phonon interaction. If this is the main
pairing interaction the gap would be largest on the
nested Fermi surface sheets. This type of interaction favors a singlet
state and unless there is an additional repulsive interaction,
such as spin-fluctuations or a strong Coulomb repulsion,
will not lead to a state with any sign changes in the order parameter.
Spin fluctuations associated with nearness to an SDW resulting
from the nested 1D sheets could lead to a specific heat enhancement,
i.e. a larger $\lambda_{tot}$, as discussed by Jiao and co-workers.
\cite{jiao}
As mentioned, this may not be needed considering the value of the
bare $N(E_F)$ from our calculations.
In any case, such an interaction cannot lead to superconductivity on
the nested sheets with the Fermi surface structure of this compound.
The reason is that in a singlet channel, spin fluctuations are
repulsive, which would favor an order parameter that changes sign
between the nested sheets on opposite sides of the zone center, i.e.
a triplet state, while in a triplet channel spin fluctuations are
attractive, which would favor an order parameter that is the same
on opposite sides of the zone center, i.e. not a triplet.
Besides the possibility of pairing due to nearness to a CDW, it is
to be noted that there are other layered tellurides, in particular
Ir$_{1-x}$Pt$_x$Te$_2$ that become superconducting 
\cite{pyon}
and which do not have Fermi surface nesting features that
lead to phonon softening.
\cite{cao}
Thus it may well be that Ta$_4$Pd$_3$Te$_{16}$ is also an s-wave
superconductor with pairing from phonons associated with the
Te-Te $p$ bonding and a Fermi surface associated Te $p$ bands.

Recently, Pan and co-workers reported thermal conductivity measurements
on a single crystal sample with a low residual resistivity below 
4 $\mu$$\Omega$ cm.
They found a large linear electronic contribution persisting down to
$T$=80 mK, and furthermore that this contribution increases with
field similar to the behavior of a cuprate superconductor.
This implies the presence of ungapped parts of the Fermi surface,
such as line nodes. The data are reminiscent of cuprates and also certain of the
Fe-based superconductors, for which there is other evidence
of line nodes. \cite{hashimoto}
The observed behavior is inconsistent with the behavior of the two gap
superconductor NbSe$_2$. On the other hand, it could be compatible
with a clean multigap superconductor where the gap ratio is larger,
and in fact that data has some similarity to MgB$_2$,
\cite{sologubenko}
which is an s-wave electron phonon superconductor.
Considering the magnitude of the linear thermal conductivity
seen experimentally and using the Wiedemann-Franz relation,
one would have to assume that parts of  Fermi surface contributing $\sim$ 30\%
of the conductivity have very small gaps, in order to explain the
data without line nodes. This is possible considering the Fermi surface
structure, in which several different sheets contribute to the
conductivity and to $N(E_F)$.
The fact that these sheets derive from $p$ states associated
with different Te atoms makes more plausible large differences in the coupling
on different sheets, which may then lead to large
differences in the gaps for a clean sample.

It will be of interest to study samples with higher levels
of disorder in order to distinguish these two possibilities.
Specifically, with line nodes, disorder is expected to suppress
the ordering temperature, while in the multiband case disorder
would enhance the gap on the low gap parts of the Fermi surface
and thereby suppress the electronic thermal conductivity at low temperature.
Also, we note that the electronic thermal conductivity has the same
anisotropy as the electronic conductivity from the
ungapped parts of the Fermi surface. We find sheets with
very different anisotropies, 1D sheets, a 2D sheet and a 3D sheet.
Therefore, in a scenario in which there is a small gap on some
part, the particular sheets involved in the electronic
thermal conductivity (i.e. the low gap sheets) could be identified from the
anisotropy of the electronic part of the thermal conductivity in the
superconducting state.

To summarize, electronic structure calculations for 
Ta$_4$Pd$_3$Te$_{16}$ show four sheets of Fermi surface
derived primarily from Te $p$ states. Importantly, the transition
metal contribution to the density of states is too low to place the
system near magnetism.
The calculated value of $N(E_F)$=9.6 eV$^{-1}$ is higher than
that from a prior calculation \cite{alemany}, which can resolve
a discrepancy between the specific heat jump and the specific
heat renormalization. The Fermi surface includes
nested 1D-like sections, a 2D-like section and a heavy
3D section, which makes the largest contribution to $N(E_F)$.
The net result is a rather anisotropic but 3D metal.
Thus Ta$_4$Pd$_3$Te$_{16}$ is a multiband superconductor with
an electronic structure derived mostly from Te $p$ states.

\acknowledgments

This work was supported by the U.S. Department of Energy,
Basic Energy Sciences, Materials Sciences and Engineering Division.

\bibliography{Ta4Pd3Te16}

\end{document}